\documentclass[12pt]{article}
\usepackage{amsmath}
\usepackage{amsfonts}
\usepackage[OT1]{fontenc}
\usepackage[applemac]{inputenc}
\usepackage{mltex}

\usepackage[french]{babel}
\newtheorem{lemma}{Lemma}[section]
\newtheorem{theorem}[lemma]{Theorem}
\newtheorem{proposition}[lemma]{Proposition}

\newtheorem{remark}[lemma]{Remark}

\newtheorem{definition}[lemma]{Definition}

\def\sq{\hbox {\rlap{$\sqcap$}$\sqcup$}}
\def\sq{\hbox {\rlap{$\sqcap$}$\sqcup$}}
\def\R{ {\rm R \kern -.31cm I \kern .15cm}}
\def\C{ {\rm C \kern -.15cm \vrule width.5pt \kern .12cm}}
\def\1{{\rm 1\mskip-4.5mu l} }
\def\lsim{\raise0.3ex\hbox{$<$\kern-0.75em\raise-1.1ex\hbox{$\sim$}}}
\def\gsim{\raise0.3ex\hbox{$>$\kern-0.75em\raise-1.1ex\hbox{$\sim$}}}

\def\beq{\begin{equation}}   \def\eeq{\end{equation}}
\def\bea{\begin{eqnarray}}  \def\eea{\end{eqnarray}}

\newcommand\mysection{\setcounter{equation}{0}\section}
\renewcommand{\theequation}{\thesection.\arabic{equation}}
\newcounter{hran} \renewcommand{\thehran}{\thesection.\arabic{hran}}

\def\bmini{\setcounter{hran}{\value{equation}}
    \refstepcounter{hran}\setcounter{equation}{0}
    \renewcommand{\theequation}{\thehran\alph{equation}}\begin{eqnarray}}

\def\bminiG#1{\setcounter{hran}{\value{equation}}
\refstepcounter{hran}\setcounter{equation}{-1}
\renewcommand{\theequation}{\thehran\alph{equation}}
\refstepcounter{equation}\label{#1}\begin{eqnarray}}

\def\emini{\end{eqnarray}\relax\setcounter{equation}{\value{hran}}\renewcommand{
\theequation}
{\thesection.\arabic{equation}}}

\begin{document}
%\begin {center}
\title {About Quantum Revivals, Quantum Fidelity\\
A Semiclassical Approach}
%\end {center}
\author{\it {\bf Monique Combescure} \\
\it IPNL, B\^atiment Paul Dirac \\
\it 4 rue Enrico Fermi,Universit\'e Lyon-1 \\
\it  F.69622 VILLEURBANNE Cedex, France\\
\it email monique.combescure@ipnl.in2p3.fr}
\vskip 1 truecm
\date{}
\maketitle

\begin{abstract}
The aim of this paper is three-fold. First to establish in a clear and rigorous way a formula
proposed heuristically by Mehlig-Wilkinson for the Metaplectic Operators corresponding to
a given Symplectic Transformation in classical Phase-Space.
\\
Second this formula is applied to the study of Quantum Recurrences, which has attacted great 
interest in recent years (see \cite{ro} for a complete account of the recent approaches).
 The return probability is given by the
 squared modulus of the overlap between a given initial wavepacket and the corresponding 
 evolved one; quantum recurrences in time can be observed if this overlap is unity. We provide
  some conditions under which this is semiclassically achieved taking as initial wavepacket a 
  coherent state localized on a closed orbit of the corresponding classical motion.\\
Third, we start a rigorous approach of the
 ``quantum fidelity'' (or Loschmidt Echo): it is the squared modulus of the overlap of an 
evolved quantum state with the same state evolved by a slightly perturbed Hamiltonian. It has
 attracted a great interest in the last decade, for the purpose of ``Quantum Chaos'' problems,
 and to quantum computation analysis. However the results are most of the time not entirely 
 conclusive, and even sometimes  contradictory. Thus it is useful to start a rigorous
 approach of this object.\\
  The decrease in time of the Quantum Fidelity measures the sensitivity of quantum evolution
  with respect to small 
 perturbations.  Starting with suitable initial quantum 
 states, we develop a semiclassical estimate of this quantum fidelity in the Linear Response
  framework (appropriate for the small perturbation regime), assuming some ergodicity
   conditions on the corresponding classical motion.

\end{abstract}
\newpage

\mysection{Introduction}
The revival of quantum wavepackets along the time evolution has attracted much recent interest (see \cite{ro} and references therein contained), and we want to stress here 
how much can be demonstrated about this topic in the semiclassical limit, starting from a wavepacket (coherent state) conveniently localized around a point on a closed classical 
orbit (Section 3) It consists in considering the overlap of a time evolved quantum wavepacket with its initial state at conveniently chosen times called ``revival times''.\\

 In addition a large physical literature has reported (essentially heuristic) results about the so-called ``quantum fidelity'' (also called Loschmidt echo) in the perturbative and/or
  semiclassical limit (see \cite{beca1},\cite{beca2}, \cite{becave1}, \cite{becave2}, \cite{ceto1}, \cite{ceto2},\cite{cupawi},\cite{cupaja},\cite{cudapazu},
  \cite{emwellco},
 \cite{fihe},\cite{gillma},\cite{goprose},\cite{jaadbe1},\cite{jaadbe2},
 \cite{japa2},\cite{pe},\cite{pro},\cite{prose},\cite{prosezni}, \cite{prozni},\cite{sala},\cite{schlunk}\cite{sitwobe},\cite{vaco},\cite{vahe1}, 
 \cite{vahe2},\cite{vepro},\\
 \cite{wang1},\cite{wang2}, \cite{wellco},\cite{weemllco},\cite{welltsa},\cite{wi},\cite{wico},\cite{znipro}). The idea, which goes back to Peres \cite{pe}
 is to study the behavior in time of 
 the sensitivity of quantum evolutions with respect to small perturbations of the Hamiltonian, mainly in situations where the underlying classical motion is ``chaotic''.\\
  More precisely, one considers the overlap of the perturbed quantum evolution of some given initial wavepackets with that under the unperturbed quantum dynamics, in 
  particular the decay properties in time of this overlap.\\
  
   Here we shall consider exact results about this behavior in the ``linear response'' framework, starting with a bunch of
   initial wavepackets taken as eigenstates of the unperturbed Hamiltonian, assuming some ergodicity or mixing assumptions about the underlying classical flow, and taking the
    semiclassical limit $\hbar \to 0$, (Section 3).\\
  
  We note that these results constitute a first step towards a completely rigorous understanding  of these topics in the semiclassical regime.\\
  
  The paper is organized as follows. In section 2, we give a proof of a beautiful formula proposed by Mehlig-Wilkinson \cite{mewi} for the metaplectic operators associated to a given 
  symplectic map. (This topic is being developed and completed in a recent work in collaboration with D. Robert). This formula appears to be very useful in the semiclassical
   study of the Quantum Revivals that we develop in Section 3. Section 4 contains some results about the revivals of coherent states in the case of time-periodic Hamiltonians 
   that are {\bf quadratic} in coordinates and momenta. In Section 5 we give some result about the Quantum Fidelity in the semiclassical regime, in the Linear-Response
    framework, and recover a result derived heuristically by Prosen and Znidaric \cite{prozni}. Section 6 is an appendix containing intermediate lemmas necessary for the
    proof of Theorem 3.6. In Section 7 we give some concluding remarks.
  \newpage
  \mysection{A proof of the Mehlig-Wilkinson Formula}
  
  In a recent paper, Mehlig and Wilkinson proposed (heuristically) a beautiful formula expressing the metaplectic operators corresponding to suitable symplectic matrices $M$
  not having 1 as eigenvalue. (\cite{mewi}). In a paper in preparation with D. Robert (\cite{coro3}), we give a developed and generalized proof of this formula, 
   including the calculus of the phase, and the case where the Symplectic Matrix has 1 as eigenvalue. Here we present a simple proof, closer to the physical intuition, in the same
    spirit as in \cite{coro3} but without expliciting the phase factor.\\
  
  A metaplectic transformation implements in the quantum world the symplectic transformations $Sp(2n)$ (canonical transformations) of classical mechanics.
   To a given a symplectic transformation $M$, it corresponds unitary operators $\hat R(M)$ in the Hilbert space of quantum states acting covariantly on the canonical operators 
   of position and momentum $\hat Q ,\quad \hat P$ in the following way:
  
  \beq
  \label{1.1}
  \hat R( M)^*\left(
  \begin{array}{c}
  \hat Q\\
  \hat P
  \end{array}
 \right )
  \hat R(M) = M \left(
  \begin{array}{c}
  \hat Q\\
  \hat P
  \end{array}
  \right)
 \eeq
 \noindent 
  plus some algebraic conditions ensuring that $M \mapsto \hat R(M)$ defines a projective representation of $Sp(2n)$, with sign indetermination. For any $M \in Sp(2n)$ we
   can find a $C^1$-smooth curve $F_{t}, \ t \in [0,1]$ in $Sp(2n)$ such that $F_{0}= \1$ and $F_{1}= M$. $F_{t}$ is clearly the linear flow defined by the quadratic 
   Hamiltonian $H_{t}(z):= \frac{1}{2}z.S_{t}z$ with
  
  \beq
  \label{1.2}
  S_{t}= - J\dot F_{t}F_{t}
  \eeq
  where
  \beq
  \label{1.3}
  J=\left(
\begin{array}{cc}
0&\1\\
-\1&0
\end{array}
\right)
\eeq
    If $U_{t}$ is the quantum propagator associated with $\hat H_{t}$, we define
  
  $$
  \hat R(M)= U_{1}$$
  
 which obeys equ.(\ref{1.1}). If two sympectic paths $F_{t}$ and $F_{t}'$ join 1 to $M$, then we have $U_{1}= \pm U'_{1}$.
  
  \begin{definition}
 Let $z:=(q,p)$ be a generic point in classical phase space $Z:= \mathbb R^{2n}$. The covariant symbol of an operator $\hat A$ is a function $A^{\sharp}(z)$ defined in
  the whole phase-space $\mathbb R^{2n}$ such that
  \beq
  \label{1.4}
  \hat A = \int dz A^{\sharp}(z)\hat T(z)
  \eeq
  with $\hat T(z)$ being the Weyl Heisenberg operator
  \beq
  \label{1.5}
  \hat T(z):= \exp\left(\frac {ip.\hat Q- iq. \hat P}{\hbar}\right)
  \eeq
  
  \end{definition}
  
  \begin{theorem}
  Assume that $M$ doesn't have 1 as eigenvalue. Then the covariant symbol of $\hat R(M)$ is given by
  \beq
  \label{1.6}
  \Omega(z)= \frac{\gamma_{M}}{\vert \det(\1 - M)\vert^{1/2}}\exp\left(\frac{i}{2\hbar}z.Az\right)
  \eeq
  where $\gamma_{M}$ is a complex number of modulus 1, and $A$ is defined by
  \beq
  \label{1.7}
  A := \frac{J}{2}( M- \1)^{-1}(\1 +M)
  \eeq
  and $J$ is the symplectic matrix defined in equ. (\ref{1.3}).
 
  \end{theorem}
  
    Proof: By the inversion formula we get
  \beq
  \label{1.9}
  \Omega(X) = {\rm Tr}(U_{1}\hat T(-X))
  \eeq
  \noindent
  which using the coherent states basis 
  \beq
  \label{1.10}
  \varphi_{z} := \hat T(z)\varphi_{0}
  \eeq
  where
   \beq
   \label{1.11}
  \varphi_{0} := (\pi \hbar)^{-n/4} \exp (-x^2/2\hbar)
  \eeq
can be rewritten as:
\beq
\label{1.12}
\Omega(X)= (2\pi \hbar)^{-n}\int dz \langle \varphi_{z}, \hat T(-X)U_{1}\varphi_{z}\rangle
\eeq
In all that follows we shall omit for simplicity the factors $2\pi$ and $\hbar$, that we can easily reintroduce in the result. We have:
\beq
\label{1.13}
\hat T(X)\hat T(z) = e^{-i\sigma(X,z)/2}\hat T(X+z)
\eeq
with the notation
$$\sigma(z,z') = z.Jz'$$

Clearly
\beq
\label{1.14}
U_{1}\varphi_{z}= \hat T(z')\hat R(M)\varphi_{0}
\eeq
with $z' := Mz$, so that
\beq
\label{1.15}
\Omega(X)= \int dz e^{i\sigma(X, z)/2- i \sigma(z', X+z)/2}\langle \varphi_{Y}, \hat R(M)\varphi_{0}\rangle
\eeq  
where by $Y$ we denote the variable
\beq
\label{1.16}
Y := X + z - z' = X + (\1-M)z
\eeq
The scalar product $\langle \varphi_{Y}, \hat R(M)\varphi_{0}\rangle$
 is equal to
 $$e^{iY.QY/2}$$
 with $Q$ a matrix that we do not need to make explicit.
  Performing the change of variable $z \mapsto Y$, we rewrite $\sigma(z+z', X+z)$ in terms of $X$ and $Y$ alone. The cross-terms vanish, so that we are simply left with
  \beq
  \label{1.17}
 - \sigma(z+z', X+z) = X.AX + Y.Q'Y
  \eeq
  with $Q'$ a matrix that we do not need to make explicit. For the calculus of $A$, we easily check that the quadratic terms in X are:
  \beq
  \label{1.18}
 - (\1 +M)(\1-M)^{-1}X. JM(\1-M)^{-1}X =- (\1-M)^{-1}X. J(\1+M)(\1-M)^{-1}X
  \eeq
  where we have used that, by the symplecticity of $M$, 
  \beq
  \label{1.19}
  (\1+\tilde M)J M = JM+J= J(\1+M)
  \eeq
  where $\tilde M$ denotes the transpose of $M$.  
  Therefore (\ref{1.16}) is nothing but
  \beq
  \label{1.20}
 - (\1-M)^{-1}X.JM(\1 -M)^{-1}X=- (\1-M)^{-1}X. \frac{JM -\tilde MJ}{2}(\1-M)^{-1}X
  \eeq
  However, using again the symplecticity of $M$, we have:
  $$(\1-\tilde M)^{-1}(JM- \tilde MJ)= J(\1 +M)$$
  Therefore (\ref{1.20}) is nothing but
  $$-\frac{1}{2}X. J(\1 +M)(\1-M)^{-1} X \equiv X.AX$$
  
  Therefore inserting this result into (\ref{1.15}) simply leads to:
  \beq
  \label{1.21}
  \Omega(X)= \Omega(0)e^{iX.AX/2}
  \eeq
  where $\Omega(0)$ is the integral over $Y$ that we do not perform exactly, although it is pure Gaussian integral, but we shall content ourselves to determine its absolute value.
  
  \medskip
  
 \begin{lemma}
 We have
 \beq
 \label{1.22}
 \vert \Omega(0) \vert = \vert \det (\1-M)\vert^{-1/2}
 \eeq
 \end{lemma}
 
 \medskip
  
 Proof:
 $$\langle \varphi_{z}, U_{1}\varphi_{z}\rangle = \int dX U_{1}(X) W_{\varphi_{z}}(X)$$
 where $U_{1}(X)$ is the Weyl symbol of the quantum evolution operator $U_{1}$, and $W_{\varphi_{z}}(X) = W_{\varphi_{0}}(X-z) =  \Phi_{0}(X-z)$
 with
 \beq
 \label{1.23}
 \Phi_{0}(z) := e^{-z^2}
 \eeq
 being the Wigner function of $\varphi_{0}$. Thus
 \beq
 \label{1.24}
 \langle \varphi_{z}, U_{1}\varphi_{z}\rangle = U_{1}* \Phi_{0}(z)
 \eeq 
  Let us now calculate $\int dz \vert \langle \varphi_{z}, U_{1}\varphi_{z}\rangle \vert^2$ by using Plancherel Theorem:
  \beq
  \label{1.25}
  \int dz \vert \langle \varphi_{z}, U_{1}\varphi_{z}\rangle \vert^2 = \int dz \vert (U_{1}* \Phi_{0})(z)\vert ^2
  \eeq
  $$= \int dz \vert\mathcal F(U_{1}* \Phi_{0})(z)\vert^2$$
   where $\mathcal F $ denotes the symplectic Fourier Transform. But $\Omega(z)$ is  the symplectic Fourier Transform of $U_{1}(z)$. Thus equ. (\ref {1.25}) becomes
 \beq
 \label{1.26}
 = \int dz \vert \Omega(z)\vert^2 \vert \mathcal F \Phi_{0}(z)\vert^2 = \vert \Omega(0)\vert^2 \int dz e^{-z^2/2} = \vert \Omega(0)\vert ^2
 \eeq
 because $A$ being real symmetric (which can be deduced easily from the symplecticity of $M$), $\vert\Omega(z)\vert \equiv \vert \Omega(0)\vert , \quad 
 \forall z\in \mathbb R^{2n}$. \\

 Let us now calculate the same quantity using the Wigner Transforms:
 \beq
 \label{1.27}
 \vert \langle \varphi_{z}, U_{1}\varphi_{z }\rangle \vert ^2 = 2\int_{\mathbb R^{2n}} du\  W_{\hat T(z)\varphi_{0}}(u)\ W_{\hat T(z')\hat R(M)
 \varphi_{0}}(u)
 \eeq 
 $$= 2\int du W_{\varphi_{0}}(u-z) \ W_{\hat R(M)\varphi_{0}}(u-z')= 2\int du \exp\left[- (u-z)^2 - (M^{-1}u - z)^2\right]$$
 Now, in order to integrate equ. (\ref{1.27}) over $z$, we perform the change of variable
 $$z \mapsto X := z-u$$ so that we are left with a simple Gaussian of the form
 \beq
 \label{1.28}
 2\int dX du\  e^{-X^2 - (X- Y)^2}= \int du\  e^{-Y^2/2}= \vert \det(\1 - M^{-1})\vert^{-1}
 \eeq
  since $Y := (M^{-1}- \1)u$. But since $M$ is symplectic, it has determinant one and thus equ. (\ref{1.28}) equals $$\vert \det (M - \1)\vert^{-1}$$
   Hence,
  \beq
  \label{1.29}
  \vert \Omega(0)\vert^2 \equiv \vert \det (\1 -M)\vert ^{-1}
  \eeq
  \\
  \sq
  
  Theorem 2.2 is nothing but the beautiful Mehlig-Wilkinson formula (\cite{mewi}) for the metaplectic representation. In \cite{coro3} , we establish it in full generality,
   including the calculus of the phase factor $\gamma (F)$. Assume that $F$ is a symplectic map not having 1 as eigenvalue. Then:
  \beq
\label{2.25}
\hat R(F)= h^{-n}\frac {\gamma(F)}{\vert \det (\1 - F)\vert^{1/2}}\int_{\mathbb R^{2n}}dz \exp \left(\frac {iz.Az}{2\hbar}\right)\hat T(z)
\eeq
where $\gamma(F)$ is a complex number of modulus 1 that we do not specify, and A is the $2n \times 2n$ real symmetric matrix
\beq
\label{2.26}
A:= \frac {1}{2}J(F + \1)( F-\1)^{-1}
\eeq
and J is the symplectic matrix (\ref{1.3})

  \newpage
  \mysection{Quantum Revivals}
  We consider a quantum wavepachet $\varphi \in \mathcal H = L^2(\mathbb R^n)$, together with a quantum Hamiltonian $\hat H$ selfadjoint in $\mathcal H$.
   According to 
  Schrödinger equation, the quantum evolution of wavepackets is provided by the unitary group of evolution $U_{H}(t):= \exp(-it \hat H/\hbar)$, $\hbar$ being the Planck 
  constant (divided by $2\pi$). The quantum revivals of that wavepacket are observed by considering the overlap between the evolved wavepacket $U_{H}(t)\varphi$ with 
  the initial one $\varphi$:
  \beq
  \label{2.1}
  R(t):= \langle \varphi, U_{H}(t)\varphi \rangle
  \eeq
  
  Assume now that the initial wavepacket is taken as a Gaussian one, namely a coherent state 
  $\varphi_{z}$ defined in equ. (2.9-10) for any $z := (q,p)\in Z 
  :=\mathbb R^{2n}$.\\
  
The coherent states (sometimes written in the Dirac notation $\vert z \rangle$ instead of $\varphi_{z}$) have been shown to display remarkable semiclassical 
propagation properties (see \cite{coro1}) that we recall  here:
\\

Along a tradition which goes back to Hepp (1974), one can start by ``following'' a coherent state along its semiclassical evolution. We shall establish the following result:\\
starting from a coherent state $\vert z \rangle$ at time t=0, its quantum evolution stays, up to a phase, close to a ``squeezed state'' $$\hat T(z_{t})\hat R(F_{t})\vert 
0 \rangle$$
centered around the point $z_{t}:= \phi_{H}^t(z)$, ($\Phi_{H}^t$ being the classical flow generated by the classical symbol $H$ of $‹\hat H$),
 with a ``dispersion'' governed by a symplectic matrix $F_{t}$ that we shall make precise later. This approximation is of 
order $O(\hbar^{\epsilon})$ as long as time doesn't go beyond the so-called {\bf Ehrenfest time} $T_{E}:= \lambda^{-1}\log \hbar ^{-1}$. Intuitively the phase-space
 directions where the wavepacket {\bf spreads} are the unstable directions of the classical flow, whereas those along which they {\bf are squeezed} are the stable ones; 
 those {\bf stable} and {\bf unstable} directions are encoded in the symplectic matrix $F_{t}$.

\bigskip
All that follows will be true for a very general class of Hamiltonians (possibly time-dependent):
\beq
\label{2.5}
 \exists m, M, K >0: \quad (1+ z^2)^{-M/2}\vert \partial_{z}^{\gamma}H(z, t)\vert \leq K \quad \forall \vert \gamma \vert \geq m
\eeq

uniformly for $(t, z)\in [-T, T]\times Z$
\\

such that the classical and quantum evolutions respectively (for the classical symbol and its Weyl quantization resp.) exist for $t \in [-T, T]$.

\bigskip
It is well-known that the stability of the classical Hamiltonian
evolution governed by $H(z, t)$ is given by the following linear system:
\beq
\label{2.6}
\dot F = JM_{t}F
\eeq
where $M_{t}$ is the $2n \times 2n$ Hessian matrix of H at point $z_{t}$ of the classical trajectory:
\beq
\label{2.7}
(M_{t})_{j,k}:=\left(\frac {\partial^{2}H}{\partial z_{j}\partial z_{k}}\right )_{j, k}(z_{t}, t)
\eeq
is symmetric real,
 J is the symplectic matrix (\ref{1.3}) and the initial datum is 
\beq
\label{2.9}
 F(0) \equiv \1
\eeq

\medskip
Consider the purely quadratic Hamiltonian (time-dependent):
\beq
\label{2.10}
\hat H_{0}(t):= \frac {1}{2}\hat Z. M_{t}\hat Z
\eeq
It induces a quantum evolution $U_{0}(t, t')$ via
\beq
\label{2.11}
i\hbar \frac {\partial}{\partial t}U_{0}(t, t' )= \hat H_{0}(t)U_{0}(t, t')
\eeq
which is entirely explicit:

\begin{lemma}
Let $F_{t}$ be the $2n \times 2n$ symplectic matrix solution of (\ref{2.6})-(\ref{2.9}). We note by $\hat R (F_{t})$ the associated metaplectic operator, unitary in
 $\mathcal H$. We have:
\beq
\label{2.12}
U_{0}(t, 0) = \hat R(F_{t})
\eeq
and therefore by the chain rule:
\beq
\label{2.13}
U_{0}(t, t')= \hat R (F_{t})\hat R (F_{t'}^{-1})
\eeq

\end{lemma}

The proof of this result is classical and can be found in \cite{per}.

 It encompasses the physical intuition that, for Hamiltonians purely {\bf quadratic},
 the quantum dynamics is exactly solvable in terms of the classical one. Namely the linear equ. (\ref{2.6}) is nothing but the classical Hamilton's equations for the quadratic 
 Hamiltonian (\ref{2.10}).

\medskip
In fact $\hat R (F_{t})$ decomposes itself into the product of two unitaries,(built from the symplectic matrix $F_{t}$)\\
-one expressing the ``squeezing''\\
-one expressing the ``rotation''

\begin{lemma}
We define the usual ``creation and annihilation'' operators of quantum mechanics as follows:
\beq
\label{2.14}
a := \frac {\hat Q + i \hat P}{\sqrt{2\hbar}} \qquad a^{\dagger}:= a^*= \frac {\hat Q - i \hat P}{\sqrt{ 2\hbar}}
\eeq
From $F_{t}$ can be built two $2n \times 2n$ matrices $E_{t}$ and $\Gamma_{t}$ such that:
\beq
\label{2.15}
\hat R (F_{t})= \hat S (E_{t}) \hat  R (t)
\eeq

\beq
\label{2.16} 
\hat S(E_{t}): =\exp\left(\frac {1}{2}(a^{\dagger}. E_{t}a^{\dagger} - a. E_{t}^*a)\right)
\eeq
\beq
\label{2.17}
\hat  R (t)= \exp \left( \frac {i}{2}(a^{\dagger} . \tilde\Gamma_{t} a + a.\Gamma_{t} a^{\dagger})\right)
\eeq
\end{lemma}

\medskip
Morally, in dimension n=1, $\hat  R (t)$ has the simple form
$$\exp \left(\frac {i\gamma_{t}}{2}(\hat Q^2 + \hat P^2)\right)$$
which is simply a rotation by the real angle $\gamma_{t}$.

\bigskip
Let us consider the Taylor expansion up to order 2 of Hamiltonian $H(z, t)$ around the point $z= z_{t}:= (q_{t}, p_{t})$ of the classical trajectory at time t:

\beq
\label{2.18}
H_{2}(t):= H(z_{t}, t) + (z-z_{t}). \nabla H(z_{t}, t) + \frac {1}{2}(z-z_{t}). M_{t}(z-z_{t})
\eeq
By quantization it yields:
\beq
\label{2.19}
\hat H_{2}(t)= H(z_{t}, t)\1 + (\hat Z-z_{t}).\nabla H(z_{t}, t) + \frac {1}{2}(\hat Z- z_{t} ). M_{t}(\hat Z -z_{t})
\eeq

\medskip
Let $U_{2}(t, s)$ be the quantum propagator for Hamiltonian $\hat H_{2}(t)$. We have:

\begin{proposition}
\beq
\label{2.20}
U_{2}(t,s)= e^{i(\delta_{t}- \delta_{s})/\hbar}\hat T(z_{t})\hat R (F_{t})\hat R(F_{s}^{-1})\hat T(-z_{s})
\eeq
where
 \beq
 \label{2.21}
\delta_{t}:= S_{t}(z) - \frac {q_{t}.p_{t}-q.p}{2}
\eeq
and $S_{t}(z)= \int ds (\dot q_{s}.p_{s} - H(z_{s}, s))$ is the classical action along the trajectory $z\to z_{t}$
\end {proposition}

The proof of this result can be found in \cite{coro1}.

\bigskip
In fact this propagator which, being constucted via generators of the coherent/squeezed states acts in a simple manner on coherent states, and appears to be a good 
approximation of the full propagator $U_{H}(t, s)$ in the classical limit, when acting on coherent states:

$$U_{2}(t,0)= e^{i\delta_{t}/\hbar}\hat T(z_{t})\hat S(E_{t})\hat  R(t)\hat T(-z)$$
$$U_{2}(t, 0)\vert z \rangle = e^{i\delta_{t}/\hbar}\hat T(z_{t})\hat S(E_{t})\hat  R(t)\vert 0 \rangle$$
$$=e^{i\delta_{t}/\hbar + \gamma_{t}}\hat T(z_{t})\hat S(E_{t})\vert 0 \rangle$$
where $\gamma_{t}= \frac {1}{2} tr \Gamma_{t}$\\
The state 
\beq
\label{2.22}
\Phi (z, t):= \hat T(z_{t})\hat S(E_{t})\vert 0 \rangle
\eeq
is simply a squeezed state centered in $z_{t}$, with a ``squeezing'' given by matrix $E_{t}$.

\bigskip
\begin{theorem}
Let $H$ be an Hamiltonian satisfying the assumptions (\ref{2.5}) and the existence of classical and quantum flows for $t \in [-T, T]$. Then we have, uniformly for $(t, z) 
\in [-T, T]\times Z$:
\beq
\label{2.23} 
\Vert U_{H}(t, 0)\varphi_{z}- e^{i \delta_{t}/\hbar + \gamma_{t}}\Phi (z, t)\Vert \leq C \mu(z, t)^P\vert t \vert \sqrt \hbar\theta(z, t)^3
\eeq
P being a constant only depending on M and m, and
$$\mu (z, t):= Sup_{0\leq s \leq t}(1 + \vert z_{s}\vert)$$
$$\theta (z, t):= Sup_{0 \leq s \leq t}(tr F_{s}^*F_{s})^{1/2}$$
\end {theorem}

The proof of this result is contained in \cite{coro1}.

\bigskip
The estimate (\ref{2.23}) contains the dependence in $t, \hbar , z$ of the semiclassical error term. One hopes that this error remains small  when $\hbar \to 0$,
 provided that $z$ belongs to some compact set of phase-space, and $\vert t \vert$ is not too large.

\medskip
Typically
$$\theta (z, t)\simeq e^{ t\lambda} $$ where $\lambda$ is some Lyapunov exponent that expresses the ``classical instability'' near the classical trajectory. The RHS of
 equ. (\ref{2.23}) is therefore $O (\hbar^{\epsilon/2})$ provided
\beq
\label{2.24}
\vert t \vert < \frac {1-\epsilon}{6\lambda}\log \hbar ^{-1}
\eeq
which is typically the Ehrenfest time, up to a factor 1/6 that is probably inessential.

\begin{remark}
(see \cite{coro1})
Theorem (3.4) can be modified (and therefore also the state $\Phi (z, t)$) to have an estimate in $$\mu(z,t)^{lP}\sum _{j=1}^l \left( \frac {\vert t \vert}
{\hbar}\right)^j(\sqrt \hbar \theta(z,t))^{2j + l}$$
and therefore typically $O( \hbar^{l/2})$ with l integer as large as one wants. The squeezed state however now depends on l, and is typically a finite linear combination
 of wavepackets of the form:
$$\hat T(z_{t})\hat R(F_{t})\vert \Psi_{\mu}\rangle$$ where the $\Psi_{\mu}$ are excited levels of the Harmonic Oscillator in dimension n.
\end{remark}

We shall now make use of the Mehlig-Wilkinson formula (\ref{2.25}), in order to calculate the dominant
 contribution to $R(t)$, the quantum overlap at time t .

Of course this formula holds true as long as 1 is not an eigenvalue of F, but it can be generalized to this case also (see \cite{coro3}).

\bigskip
We now rewrite equ. (\ref{2.23}) as:
\beq
U_{H}(t, 0)\vert z \rangle = e^{i\delta_{t}/\hbar}\hat T(z_{t})\hat R (F_{t})\vert 0 \rangle + \varepsilon(t, \hbar, z)
\eeq
%so that:
%\beq
%R(t)= \frac {h^{-n}e^{i\delta_{t}/\hbar + \gamma_{F}}}{\vert \det(\1 - F_{t})\vert ^{1/2}}\int_{Z} dz' \langle z \vert \hat T(z_{t})\hat T(z')\vert 0 \rangle e^{i z' . A z'/2\hbar} + \varepsilon(t, \hbar, z)
%\eeq
%\beq
%= \frac {h^{-n}e^{i\delta_{t}/\hbar + \gamma_{F}}}{\vert \det(\1 - F_{t})\vert ^{1/2}}e^{i \sigma(z, z_{t})/2\hbar} \int dz' \langle 0 \vert \hat T(z_{t}-z)\hat T(z')\vert 0 \rangle  e^{i z' . A z'/2\hbar} + \varepsilon(t, \hbar, z)
%\eeq
%$$= \frac {h^{-n}e^{i\delta_{t}/\hbar + \gamma_{F}}}{\vert \det(\1 - F_{t})\vert ^{1/2}}e^{i \sigma(z, z_{t})/2\hbar} \int _{Z}dz' \exp(-i \sigma(z_{t}-z, z')/2\hbar + iz' . A z'/2\hbar- (z' + z_{t}-z)^2/4\hbar) + \varepsilon(t, \hbar, z)$$
%\beq
%= e^{i\phi} \vert \det (\1 - F_{t})\vert ^{-1/2}\vert \det (\1/2 -iA)\vert^{1/2}e^{-(z_{t}-z). B (z_{t}-z)/4\hbar} + \varepsilon(t, \hbar, z)
%\eeq
%where $\phi$ is some phase, and B is defined as follows:
%\beq
%B := \1 + (\tilde J + i\1)(\1 - 2iA)^{-1}(J + i\1)
%\eeq
%We deduce the following result:
We shall consider the following quantity (return probability):
\beq
\label{2.50}
R(\alpha,t):=\vert\langle \varphi_{\alpha},  U_{H}(t,0)\varphi_{\alpha }\rangle \vert
\eeq

\noindent
 It can be rewritten as 

\beq
\label{2.28}
\left\vert \langle \hat T(\alpha)\varphi_{0}, \ \hat T(\alpha_{t})\hat R( F_{t})
\varphi_{0}\rangle \right \vert + \varepsilon(t, \alpha, \hbar)
\eeq

The approximant (\ref{2.28}) is used to obtain the following main result of this section which is a semiclassical approximation of $R(\alpha,t)$ ( the return
probability) : Let us define the following $2n \times 2n$ matrix

\beq
\label{2.35}
K:= (-J+i\1)(\1 -2iA)^{-1}(J+i\1)
\eeq

\begin{theorem}
There exists a constant $c_{t}, \quad 0< c_{t}\le 1$ only depending upon $F(t)$ such that $R(\alpha,t)$ has the following
 semiclassical approximation:
\beq
\label{2.37}
c_{t}\left\vert \exp\left(-\frac{1}{4\hbar}(\alpha - \alpha_{t}). [\1 -(\1 +iJ)(\1 -2iA)^{-1}(\1 -iJ)](\alpha- \alpha_{t})\right)\right\vert
\eeq
\end{theorem}
The proof is rather technical, although not difficult, relying essentially on Gaussian integrations, and postponed to the Appendix.

\begin{remark}
The fact that the exponential factor in  (\ref{2.37}) is a truly decreasing factors as $\alpha - \alpha_{t}$ increases has been established in \cite{coro3}, 
using the properties of the matrix $A$ (essentially the fact that the eigenvalues of the matrix $\frac{\1}{2}-iA$ all lie on the circle $\vert z-1 \vert =1$, with $z=0
$ excluded).
\end{remark}

\bigskip
As regards the quantum revivals, we look at the expression (\ref{2.37}), and where it equals 1 when time t goes on. Clearly we need $\alpha_{t}= \alpha$, and $c_{t}= 1$.
 The first condition is realized provided $\alpha$ lies on a periodic orbit $\gamma$ of the classical flow, and $t = T_{\gamma}$ is the period of this orbit. The condition
  $c_{t}= 1$ is less transparent. It holds obviously if $F_{t}= \1,$, or more generally if $F_{T_{\gamma}}= \1 \cos \theta + J \sin \theta, \quad \theta \in [0, 2 \pi[$, 
  which is not automatically fulfilled along any periodic orbit, but can however occur for some $t= pT_{\gamma}$, $p$  being a repetition number (integer).
We thus have the following:

\begin{theorem}
Let $\gamma$ be a periodic orbit of the classical flow corresponding to the (principal) symbol of $\hat H$, wth period $T_{\gamma}$, and assume that $z \in \gamma$. 
Assume moreover that the stability matrix $F_{T_{\gamma}}$ takes the value 
\beq
\label{2.40}
F_{T_{\gamma}}=
\1 \cos \theta + J \sin \theta, \quad\theta \in [0, 2 \pi[
\eeq
 Then provided that 
$T_{\gamma}\leq \frac{1-\varepsilon}{6} \log \hbar^{-1}$, we have that the overlap $\vert L_{1}(\alpha, t)\vert =\vert \langle\varphi_{\alpha}, U_{t}
\varphi_{\alpha}\rangle \vert$ taken at value $T_{\gamma}$ obeys:
\beq
\label{2.41}
R(\alpha,t) =  1 -O(\hbar^{\varepsilon})
\eeq
\end{theorem}

This result expresses the semiclassical ``almost recurrence'' of the quantum state $U_{H}(T_{\gamma}, 0)\vert z \rangle$ to $\vert z \rangle$ for not too long periods of
 the orbit $\gamma$ on which the phase-space point $z$ sits.

\bigskip
Proof:\\
It is not hard to check that if $F_{T_{\gamma}}$ has the form above, then the matrix N given by (6.26) equals
$$N = - \frac {iJ}{2}\left(\1 (1+e^{i \theta})+ iJ (1-e^{i\theta})\right)$$
 and therefore $\vert\det N\vert = 1$, whence $c_{t} \equiv \vert \det N_{t}\vert ^{-1/2}= 1$ (see (6.30)). Note that we have no general result ensuring that it is
  actually the case that (\ref{2.40}) holds true for any
  periodic orbit $\gamma$. The only result we know for sure is that it holds true with $\theta = 0$ if the classical flow is globally periodic (see \cite{bi}). However in the 
  following section we also provide some explicit examples where the Stability Matrix can be shown explicitly to have the rotation form (\ref {2.40}).
 
 \begin{remark}
 Actually we have established (see \cite{coro3}) that $\vert \det N \vert = 1$, (and thus $c_{t}=1$) if and only if $F_{T_{\gamma}}$ is unitary, (which is a generalization of the rotation form (\ref{2.40})).
 \end{remark}
 \newpage
 
 \mysection{Quantum Revivals for some time-dependent \\
  Hamiltonians}
 
 As we have shown in Section 2, the time-dependent quadratic Hamiltonians play a special role in the understanding of Quantum Evolutions, since in particular they generate
  the Operators of the Metaplectic Group. They have the remarkable well-known property that the quantum dynamics is exactly solvable in terms of the classical dynamics. 
  They are thus a good laboratory to investigate the behavior of the stability matrix $F_{t}$ when it is propagated along a classical (closed) orbit.\\
 
 Thus in this section we investigate the Return Probability of a coherent state  for the quantum evolution $U_{t}$ generated by a quantum Hamiltonian
  of the following form:
 
 \beq
 \label{0.1}
 \hat H(t) = \frac{1}{2} \hat Z. S(t)\hat Z
 \eeq
 where $S(t)$ is a real symmetric $2n \times 2n$ matrix, and also particular non-quadratic Hamiltonians of the form
 $$\hat H(t) = \frac{\hat P^2}{2} + f(t)\frac{\hat Q^2}{2} + \frac{g^2}{\hat Q^2}$$
 
 \noindent
  In the first case, the calculus of 
 \beq
 \label{0.2}
 F_{\alpha}(t) :=\vert \langle \varphi_{\alpha}, U_{t} \varphi_{\alpha}\rangle\vert
 \eeq
 will be exact, and we make use of the result of Section 2 that the quantum propagator associated to Hamiltonian $\hat H(t)$ is exactly given by the metaplectic operator
  $\hat R(F_{t})$, where $F_{t}$ is  the stability matrix solution of equ. (3.3)
 
 \noindent
 We clearly have: (by some abuse of notation we still denote by $U_{t}(z)$ the Weyl symbol of the unitary quantum propagator $U_{t}$)
 \beq
 \label{0.4}
 \langle \varphi_{\alpha}, U_{t}\varphi_{\alpha}\rangle = \int dz W_{\varphi_{\alpha}}(z) U_{t} (z) = \int dz W_{\varphi_{0}}(z-\alpha) U_{t}(z)
 \eeq
 $$= \mathcal F ^{-1}\left( W_{\varphi_{0}^{\sharp}}(z) \Omega(z)\right)(\alpha)$$
where $\mathcal F$ is the symplectic Fourier Transform, and $$W_{\varphi_{0}}^{\sharp}(z) =( \mathcal F W_{\varphi_{0}})(z) =\langle \varphi_{0}, 
\hat T(z)\varphi_{0}\rangle\equiv e^{-z^2/4\hbar}$$
Therefore
\beq
\label{0.5}
\Omega(z)W_{\varphi_{0}}^{\sharp}(z) = \frac{\gamma_{F}}{\vert \det (\1 - F)\vert ^{-1/2}}\exp \left( -\frac{1}{2\hbar}z. (\frac{\1}{2}-iA)z\right)
\eeq
and thus
\beq
\label{0.6}
F_{\alpha}(t) = \vert \mathcal F^{-1}(W_{\varphi_{0}}^{\sharp}\Omega)(\alpha)\vert = \vert \det N_{t}\vert ^{-1/2} \exp\left\{- \frac{1}{2\hbar}J 
\alpha. B_{t}J\alpha\right\}
\eeq
where 
\beq
\label{0.7}
B_{t}:= \left(\frac{\1}{2} -iA\right)^{-1}= (F_{t}-\1)N_{t}^{-1}
\eeq
and 
\beq
\label{0.8}
N_{t} := -i \frac{J}{2}((\1 -iJ)F_{t} + \1 + iJ)
\eeq
As shown in the Appendix (Lemma 6.2), $\vert \det N_{t}\vert\ge 1$, and $B_{t} = 0$ whenever $F_{t} = \1$, so that in this case $F_{\alpha}(t)=1$.

\bigskip
Analysis of a few more or less trivial examples:\\

$$\bullet \ \hat H(t) = g(t) \frac{\hat Q. \hat P + \hat P. \hat Q}{2}$$\\

$g$ being a T-periodic function of mean zero:
$$\int_{0}^T dt g(t) = 0$$
Let $G$ be the primitive of $g$ that vanishes at $t=0$. Then clearly the stability matrix 
\beq
\label{0.9}
F_{t} = \left(
\begin{array}{cc}
e^{G(t)}& 0\\
0& e^{-G(t)}
\end{array}\right)
\eeq
is T-periodic and obeys $F_{T}=\1$ together with $N_{T} = -iJ$ so that 
$$F_{\alpha}(T) = 1$$ for {\bf any coherent state} $\varphi_{\alpha}$, namely $\varphi_{\alpha}$ is recurrent along the periodic orbits of the classical flow
 (and this is even true without any phase).\\

Note that the stability matrix $F_{t}$ is always a rotation in this case:
$$F_{t} = \left(
\begin{array}{cc}
\cos G(t) & \sin G(t)\\
-\sin G(t) & \cos G(t)
\end{array}
\right)$$

\bigskip
$$\bullet \ \hat H(t) = g(t) \frac{\hat P^2 + \hat Q^2}{2}$$\\

$g$ being again T-periodic. Then, letting $\alpha = (q,p) \in \mathbb R^2$ be any point 
in phase-space  one gets by an explicit calculus the return probability of a coherent 
state 
\beq
\label{0.10}
\langle \varphi_{\alpha}, U_{t}\varphi_{\alpha}\rangle = 
\exp\left( -\frac{1}{2}i G(t) - \frac{1}{2} i (p^2 + q^2)\sin G(t) - (p^2+ q^2)
\sin^2 \left(
\frac{ G(t)}{2}\right)\right)
\eeq
From this we deduce that if $G(t) = \int_{0}^T dt g(t)=2 k \pi,\quad \mbox{for}\  k \in \mathbb Z$, then
$$\langle \varphi_{\alpha}, U_{t} \varphi_{\alpha}\rangle = (-1)^k, \quad \forall \alpha$$

\bigskip
$$\bullet \ \hat H(t) = \frac{\hat P^2}{2} +\frac{1}{2} (\lambda \cos (\omega t )+ \mu)\hat Q^2$$\\

The classical trajectories are the solutions of Mathieu equations, which are known to be either {\bf stable} or {\bf unstable}, depending on the zones in the parameter space 
$(\lambda, \mu) \in \mathbb R^2$.\\

Assume $(\lambda, \mu)$ lie in some {\bf stability zone} of the Mathieu equation. Then there exist a real parameter $\rho$ together with an infinite sequence
 $\left \{c_{n}\right\}_{n \in \mathbb Z}$ such that the stability matrix has the following form:
\beq
\label{0.11}
F_{t} = \left(
\begin{array}{cc}
\frac{1}{C} \sum_{n}c_{n}\cos [(2n +\rho)\omega t/2]& \frac{2}{\omega D}\sum_{n}c_{n}\sin [(2n + \rho)\omega t/2]\\
  & \\
  -\frac{\omega}{2C}\sum_{n}(2n + \rho)c_{n}\sin[ (2n + \rho)\omega t/2]& \frac{1}{D}\sum_{n}c_{n}(2n + \rho)\cos [(2n +\rho)\omega t/2]
\end{array}
\right)
\eeq
where the constants $C,\ D$ are determined by
$$C:= \sum_{n} c_{n}$$
$$D := \sum_{n} (2n + \rho)c_{n}$$
In general $F_{t}$ is only {\bf quasiperiodic}, but it is known that for the limiting curves $\mu = f_{k}(\lambda)$ of the stability zones, one has $\rho = 2k, \quad k 
\in \mathbb N$, so that in this case $$F_{T} = \1, \quad T = \frac{2\pi}{\omega}$$
in which case we have exact revivals for any coherent wavepacket $\varphi_{\alpha}$.

\bigskip
Note that if the $\left\{c_{n}\right\}_{n\in\mathbb Z}$, and $\rho$ happen to satisfy
$$(2 - \omega \rho)\sum_{n \in \mathbb Z} c_{n} = 2\omega \sum_{n \in \mathbb Z}n c_{n}$$
then $F_{T}$ reduces to a simple rotation matrix
$$F_{T}= \left(
\begin{array}{cc}
\cos (\pi \rho) & \sin(\pi \rho)\\
-\cos (\pi \rho)& \cos( \pi \rho)
\end{array}
\right)$$

\bigskip
Note that it has been established in \cite{co} that beautiful recurrences can be obtained in this case in a much more general way if the initial wavepacket (coherent state)
 is chosen in an appropriate way, with a dispersion conveniently adjusted according to the classical solutions of Mathieu equations. Define
\beq
\label{0.12}
a := \frac{\omega D}{2C}
\eeq
which is a positive real number, at least in the high frequency regime (and for parameters in the stability zones). Define 
\beq
\label{0.13}
\psi(x) := (\pi \hbar)^{-1/4}a^{-1/2}\exp \left(-a \frac{x^2}{2 \hbar}\right)
\eeq
be the reference normalized wavepacket. It is the ground state of a particular Hamonic Oscillator of the form
\beq
\label{0.20}
\hat H_{0} := \frac{1}{2}\hat P^2 + \left(\frac{D \omega}{2C}\right)\frac{1}{2}\hat Q^2
\eeq
 Its recurrences are defined through the overlap

\beq
\label{0.14}
F(t):= \langle \psi, U_{t}\psi \rangle
\eeq
 It has been established in \cite{co} that one has the exact result:
 
 \begin{proposition}
 One has
 \beq
 \label{0.15}
 F(t) = e^{-i\rho \omega t/4}G(t)
 \eeq
 where $G$ is the T-periodic function defined as
 \beq
 \label{0.16}
 G(t) := \left(\sum_{n} c_{n}\left( \frac{1}{2C} + \frac{2n + \rho}{2D}\right)e^{in\omega t}\right)^{-1/2}
 \eeq
 Therefore one has revivals at times $T = \frac{2\pi}{\omega}$ up to the secular phase $\exp\left( - \frac{i \rho \omega t}{4}\right)$:
 $$\vert F(kT) \vert = 1, \quad \forall k \in \mathbb Z$$
 \end{proposition}

\begin{remark}
Note that the reference wavepacket $\psi$ is a kind of coherent (squeezed) state, localized around the fixed point $x(t) = 0$ of the Mathieu equation, which is a periodic 
orbit reduced to a point. In the paper referenced above (\cite{co}), this point corresponds to the center of a quadrupole radiofrequency trap (Paul trap), which ``traps'' an 
ion in its center, classically as well as quantum mechanically. Ideally this holds for ever, but due to imperfections of the trap, and possible occurrence of several ions trapped
 together, there is an ``escape time'' which can be evaluated with respect to the various parameters of the ion and of the trap (see \cite{co2}).
\end{remark}

\bigskip
$$\bullet \quad \hat H(t) = \frac{\hat P^2}{2} + f(t)\frac{\hat Q^2}{2} + \frac{g^2}{\hat Q^2}$$\\

 in dimension 1, $g$ being a positive constant, and $t \mapsto f$ a T-periodic function. (This Hamiltonian has been considered in \cite{co1} as a model for interacting ions 
 in a quadrupolar trap.) The important point is that the purely quadratic part of the Hamitonian governs the quantum dynamics. Let $x(t)$ be a solution of the Hill's equation
 \beq
 \label{0.17}
\ddot x(t) + f(t)x(t)=0
 \eeq
  with initial data
 \beq
 \label{0.18}
x(0)= \alpha, \quad \dot x(0) = \frac{i}{\alpha}
\eeq
 It can be shown easily that $x(t)$ can be written as
 \beq
 \label{0.19}
 x(t) := e^{u+i\theta}
 \eeq
 where $t \mapsto u, \theta$ are real functions, and $u$ is T-periodic, and that 
 $$\dot\theta = e^{-2u}$$ provided we are in a stability region of equ. (\ref{0.17}) (which are generalization of the stability zones of the Mathieu equation considered 
 above). Then it can be established: (see \cite{co1})
 
 \begin{proposition}
 Define 
 \beq
 \label{0.21}
 a := \frac{1}{2}\sqrt{1+8g^2}
 \eeq
 and for any $n \in \mathbb N$
 \beq
 \label{0.22}
 E_{n} = 2n +a +1
 \eeq
 \beq
 \label{0.23}
 \varphi_{n}(x) = \left( \frac{2n!}{\Gamma(a+n+1)}\right)^{1/2} x^{a+1/2}e^{-x^2/2}L_{n}^{a}(x^2)
 \eeq
 $L_{n}^{a}$ being the Laguerre polynomials. Then a solution of the time-dependent Schrödinger equation for $\hat H(t)$ is:
 \beq
 \label{0.24}
 \psi_{n}(x,t)= \exp\left( -i\theta E_{n}+ \frac{-u +i\dot u x^2}{2}\right)\varphi_{n}(xe^{-u})
 \eeq
 
 \end{proposition}

\bigskip
\noindent
Thus, since $u$ is T-periodic, and $\dot u (0) = 0$ due to our choice of initial data (\ref{0.18}), this result expresses the recurrence, up to a phase, of the corresponding 
set of wavepackets.

\bigskip
Proof of Proposition:\\
It is easy to check by direct computation that the unitary evolution operator corresponding to $\hat H_{t}$ is of the form $U_{t}U_{0}^*$, with
\beq
\label{0.25}
U_{t} = e^{i\dot u \hat Q^2/2}e^{iu(\hat P. \hat Q+ \hat Q. \hat P)/2}e^{-i\theta \hat H_{0}}
\eeq
where we make use of the following differential equation satisfied by $u$:
\beq
\label{0.26}
\ddot u + \dot u^2 - e^{-4u}+f=0
\eeq
\\

\sq

\mysection{Quantum Fidelity in the linear-response semiclassical regime}

Consider the following quantum Hamiltonians $\hat H_{0}$ and $\hat H := \hat H_{0} +
 \lambda V$, where $V$ is some perturbation, and $\lambda$ is a (small) 
coupling constant. We want to compare the quantum evolutions $U_{H_{0}}(t)$ under $\hat
 H_{0}$, and $U_{H}(t)$ under $\hat H$ respectively. Starting with a reference
 normalized state $\psi$, the quantum fidelity amplitude (or Loschmidt Echo), is defined as:
$$F(t):= \vert\langle U_{H_{0}}(t)\psi, U_{H}(t)\psi \rangle \vert^2$$
Of course if $t= 0$ this quantity is equal to 1, and the fidelity is assumed to decrease as time 
increases. The point is to obtain suitable regimes where something can be said
 about this decrease.
 
 \bigskip
 If we choose the reference state $\psi$ to be an eigenstate of either $\hat H_{0}$ or
  $\hat H$, then clearly {\bf  the fidelity reduces to the return
 probability} for either $\hat H$, or $\hat H_{0}$. In this sense the studies of Return 
 Probability and Quantum Fidelity are linked. 

\bigskip
As an entertainment exercise, we study the quantity $F(t)$ when $\psi$ is taken as an 
eigenstate $\psi_{j}$ of the unperturbed Hamiltonian $\hat H_{0}$ corresponding to 
discrete eigenvalue $E_{j}$, and recall an old (exact) estimate on the short time fidelity 
amplitude squared in that case.
$$ F_{j}(t) = \vert \langle \psi_{j}, U_{H}(t)\psi_{j}\rangle \vert ^2$$
which means that we only have to consider the ``return probability'' in this case.\\

The result we recall here is known as the ``Mandelstam-Tamm inequality'' \cite{mata}.
 (see also \cite{fle}):
Let
\beq
\label{3.10}
\bar E := \langle \psi, \hat H \psi \rangle
\eeq
\beq
\label{3.16}
\Delta E^2 := \langle \psi, (\hat H-\bar E)^2 \psi \rangle
\eeq

\begin{proposition}
Assume $\bar E, \  \Delta E>0$ are defined by equ. (\ref{3.10}) and (\ref{3.16}), 
respectively. Then the return probability of a state $\psi$ with respect to the quantum
 evolution $U_{H}(t)$ generated by the Hamiltonian $\hat H$, $\vert \langle \psi,
  U_{H}(t)\psi \rangle \vert^2$ remains, as long as time stays bounded above
  by $\pi \hbar / 2 \Delta E$, bounded from below by $\cos^2 (\frac {t \Delta E}
  {\hbar})$:
\beq
\label{3.22}
\vert \langle \psi, U_{H}(t)\psi \rangle\vert ^2 \geq \cos^2\left(\frac {t \Delta E}{\hbar}\right)
\eeq
\end{proposition}

\bigskip
 One of the most studied regime is the so-called ``linear response'' regime, which is a
  perturbative one, where the Dyson series of $U_{H}(t)$ is truncated up
  to second order expansion terms. Let us denote by $V(t)$ the Heisenberg observable:
\beq
\label{3.23}
V(t):= U_{H_{0}}(-t)V U_{H_{0}}(t)
\eeq
and by $\bar V(t)$ the ``mean'' of this operator on the interval $[0, t]$:
\beq
\label{3.24}
\bar V(t):= \int_{0}^{t}V(s)ds
\eeq
One defines:
\beq
\label{3.25}
f^{LR}(t):= 1 + \frac {\lambda}{i\hbar}\langle \psi, \bar V(t)\psi \rangle - 
\frac{\lambda^2}{\hbar^2}\int_{0}^{t}ds \langle\psi, V(s)\bar V(s)\psi
\rangle
\eeq
Since we are mainly interested by the absolute value (squared) of this quantity (to obtain
the ``quantum fidelity'') , we now consider $F^{LR}(t)$ to be the modulus squared of
 $f^{LR}(t)$ truncated
 to second order in $\lambda$:
\beq
\label{3.26}
F^{LR}(t)= 1 - \frac {\lambda^2}{\hbar^2}\left( 2 \Re\int_{0}^{t}ds \langle \psi,
 V(s)\bar V(s)\psi \rangle -\langle \psi,\bar V(t)\psi \rangle ^2\right)
\eeq
For such a truncated Linear-Response Fidelity, we shall perform the small Planck constant 
limit, namely the semiclassical regime under suitable assumptions on $H_{0}$:
%\beq
%\Vert U_{H_{0}}(-t)Op VU_{H_{0}}(t)- Op(V\circ \Phi_{H_{0}}^t)\Vert \leq C \hbar e^{\Gamma t}
%\eeq
%\end{proposition}

We shall now take as particular reference states the eigenstates of $\hat H_{0}$ whose 
eigenvalues lie in some neighborhood of a given classical energy $E$. 
Let us consider $\hbar$- dependent energy intervals in which the spectrum of $\hat H_{0}$
 is pure point:
$$
I(\hbar):= [\alpha(\hbar), \beta(\hbar)]
$$
\noindent
where $\alpha(\hbar)< E < \beta(\hbar)$ with $\lim_{\hbar \to 0}(\beta(\hbar)-
 \alpha(\hbar))= 0$, $\beta(\hbar)- \alpha(\hbar)\geq C \hbar$ for some $C>0$,

 and denote:
$$
\Lambda(\hbar):= \left\{j : E_{j}\in I(\hbar)\right\}
$$
Let $\psi_{j}$ be an eigenstate of $\hat H_{0}$ for some $j \in \Lambda(\hbar)$.
 For such reference state, the expression in (5.7) equals:
\beq
\label{3.27}
F^{LR}_{j}(t) = 1 - \frac {\lambda^2}{\hbar^2}\left( 2 \Re\int_{0}^t ds \int_{0}^s 
ds' \langle \psi_{j},V U_{H_{0}}(-s')V U_{H_{0}}(s')\psi_{j}\rangle - 
\langle \psi_{j},\bar V(t) \psi_{j}\rangle^2 \right)
\eeq
We make use of the following result:

\begin{proposition}
Let $\hat A$ be the Weyl quantization of a classical symbol $A$, and let us consider the 
Wigner transform associated with a given state $\psi \in \mathcal H$:
\beq
\label{3.28}
W_{\psi}(q, p):= h^{-n}\int_{\mathbb R^n}dy \bar \psi (q+\frac {y}{2})\psi (q- \frac {y}
{2})e^{ip.y/\hbar}\equiv h^{-n}\int_{Z}dz'\langle \psi, \hat T(z')
\psi\rangle e^{i\sigma(z,z')/\hbar}
\eeq
Then we have:
\beq
\label{3.29}
\langle \psi, \hat A \psi \rangle = \int_{Z}dz W_{\psi}(z)A(z)
\eeq
\end{proposition}
Let $V(z)$ (resp. $A_{s}$) be the principal symbol of $V$ (resp. $V V(s)$). Consider the 
classical flow $\Phi_{H_{0}}^t$ induced by the classical symbol of $\hat H_{0}$. 
Then we have the following semiclassical result (Egorov theorem):

\begin{proposition}
As $\hbar \to 0$ we have:
\beq
\label{3.30}
A_{s}(z) \to V(z)\ V\circ \Phi_{H_{0}}^s (z)
\eeq
uniformly for (s,z) in any compact set of $\mathbb R \times Z$.
\end{proposition}

This result goes back to Egorov \cite{eg} (see also \cite{boro})\\

Now the Wigner function of (almost any ) $\psi_{j}$, for $j \in \Lambda(\hbar)$ can also
 be shown to have very interesting semiclassical limit properties, for 
{\bf ergodic classical flows}, namely to converge towards the microcanonical measure on 
the energy surface $H_{0}= E$ as $\hbar \to 0$ in a weak sense (known as 
Schnirelman Theorem, see \cite{coro2}):

\bigskip
$H_{0}(q,p)$ being the principal symbol of $\hat H_{0}$, we denote by $\Sigma_{E}$ the
 energy surface
\beq
\label{3.31}
\Sigma_{E}:= \left\{ (q,p) \in \R^{2n}\ , \ H_{0}(q,p) = E \right\}
\eeq
and by $d\sigma_{E}$ the microcanonical measure normalized to unity:
\beq
\label{3.32}
d\sigma_{E}= \left(\int_{\Sigma_{E}}\frac {d\Sigma_{E}}{\vert\nabla H_{0}
\vert}\right)^{-1}\frac {d\Sigma_{E}}{\vert\nabla H_{0}\vert}
\eeq
where $d\Sigma_{E}$ is the Euclidean measure on $\Sigma_{E}$.

\bigskip
We make the following assumption:\\

(H) The dynamical system $(\Sigma_{E}, d\sigma_{E}, \Phi_{H_{0}}^t)$ is ergodic, 
which means:\\

for any continuous function $a$ on $\Sigma_{E}$, we have for almost every $z \in 
\Sigma_{E}$:\\
\beq
\label{3.33}
\lim_{T \to \infty}T^{-1}\int_{0}^T a \circ \Phi_{H_{0}}^t (z) dt =
 \int_{\Sigma_{E}} a(z) d\sigma_{E}(z)
\eeq

\bigskip
\begin{proposition}
Under rather technical assumptions (see \cite{coro2}) and hypothesis (H), for any 
$\hbar >0$, there exists $M(\hbar)\subseteq \Lambda(\hbar)$ depending only on 
Hamiltonian $H_{0}$ such that:
\beq
\label{3.34}
\lim_{\hbar \to 0}\left(\frac {\sharp M(\hbar)}{\sharp \Lambda(\hbar)}\right) =
 1
\eeq
and
\beq
\label{3.35}
\lim_{\hbar \to 0\ j \in M(\hbar)}\langle \psi_{j}, \hat A\psi_{j}\rangle = 
\int_{\Sigma_{E}}A(z) d\sigma_{E}(z)
\eeq
\end{proposition}

The proof of this proposition has been given in \cite{coro2}.\\

We shall employ all these results to give the semiclassical behavior of the linear-response 
quantum fidelity. 

\medskip
We define the mean and the autocorrelation function of $V$ (on the energy shell) as follows:
\beq
\label{3.36}
\bar V_{E}:= \int_{\Sigma_{E}}d\sigma_{E}(z)V(z)
\eeq
\beq
\label{3.37}
C_{V, E}(t):= \int_{\Sigma_{E}} d\sigma_{E}(z) \left(V(z) V\circ \Phi_{H_{0}}^t
 (z) - \bar V_{E}^2 \right)
\eeq

\bigskip
\begin{theorem}
Under assumption (H) we have the following result: 
\beq\label{3.38}
\lim_{\hbar \to 0}\frac {\sum_{j \in M(\hbar)}F_{j}^{LR}(t)}
{\sharp \Lambda(\hbar)} = 1 - 2\frac {\lambda^2}{\hbar^2}\int_{0}^t ds
 \int_{0}^s ds'
 C_{V, E}(s')
\eeq
\end{theorem}

The proof of this result follows easily from Propositions 5.3-5.4 applied to equ. (5.7).

\begin{remark}
This result has first appeared in \cite{prozni}, with a more heuristic approach.\\
Note that useful estimate of the second term in the RHS of (\ref{3.38}) could be obtained 
under better knowledge on the possible ``mixing properties'' of the system 
(and therefore on the decay of correlations $C_{V, E}(t)$ as t becomes large). \\
Usually physicists consider only the two first terms, claiming that it can be ``exponentiated''
 to yield an ``exponential-type'' decay of fidelity  (in the linear-response 
semiclassical regime) of the form $\exp (-C\lambda^{2} t/\hbar^2 )$, provided the
 ``mixing''
 is strong enough. $C$ is a constant given by the mixing property of the classical flow.
 The argument goes as follows:\\
 If the decay of correlation function is integrable, (for example if the mixing is exponential)
  then the integral over $s'$ in equ. (5.19) is
 bounded uniformly in $s$, and therefore the RHS of (5.19) is bounded above by
 $$1- C\frac{\lambda^2}{\hbar ^2}\vert t \vert$$
 They now ``exponentiate'' this estimate to yield the ``large-time'' behavior of
  $F_{j}^{LR}(t)$.
 We are not able to 
go further now in that direction by our rigorous approach.
\end{remark}

\section{Appendix}
We give now the proof of Theorem 3.6:
Using repeatedly the following property of the Weyl-Heisenberg operator $\hat T(z)$:

\begin{lemma}
i)
$$\hat T(z)\hat T(z'):= e^{-i\sigma(z,z')/2\hbar}\hat T(z+z')$$
ii)
$$\langle \varphi_{0},\hat T(z)\varphi_{0}\rangle= \exp(- \frac{z^2}{4\hbar})$$
\end{lemma}
we get easily that the RHS of equ. (\ref{2.28}) equals (up to the error term $\varepsilon(t, \alpha, \hbar)$):

\beq
\label{4.1}
\vert \langle \varphi_{0}, \ \hat T(\alpha_{t}-\alpha)\hat R(F_{t}) \varphi_{0}\rangle \vert
\eeq

\bigskip
Now using the Mehlig-Wilkinson formula (2.25) for $\hat R(F_{t})$, we get, provided 1 is not eigenvalue of $F_{t}$:
\beq
\label{4.3}
\vert\langle \varphi_{0}, \hat T( \alpha_{t}- \alpha)\hat R(F_{t})\varphi_{0}\rangle\vert = \vert \det (\1 - F_{t})\vert ^{-1/2}\left\vert
\int du e^{iu.Au/2\hbar}\langle \varphi_{0}, \hat T( \alpha_{t}- \alpha)\hat T(u)\varphi_{0}\rangle\right\vert
\eeq
$$= \vert \det(\1 - F_{t})\vert ^{-1/2}\left\vert\int du\langle \varphi_{0}, 
\hat T(u+\alpha_{t}- \alpha)\varphi_{0}\rangle e^{iu.Au/2\hbar}
e^{-i\sigma( \alpha_{t}-\alpha, u)/2 \hbar}\right\vert$$

We use the simplifying notation:
\beq
\label{4.2}
z_{t}:= \alpha_{t}- \alpha
\eeq

Now, forgetting the factors $\hbar$, that we shall reintroduce later by homogeneity, we get that equ. (\ref{4.1}) equals

\beq
\label{4.4}
\frac{1}{\vert \det (\1 - F_{t})\vert ^{1/2}}\left\vert\int_{\mathbb R^n} du  \exp \left( \frac{i}{2}u. Jz_{t}+ i\frac{u. Au}{2}\right)
\langle \varphi_{0}, \hat T(u+z_{t})\varphi_{0}\rangle\right\vert
\eeq
$$= \frac{1}{\vert \det (\1 - F_{t})\vert ^{1/2}}\left \vert\int du \exp 
\left(\frac{i u. Jz_{t}}{2} - \frac{1}{4}(u+z_{t})^2+ \frac{i}{2}u.Au\right)\right \vert $$\\
$$= \vert \det (\1 - F_{t})\vert^{-1/2} e^{-\frac{z_{t}^2}{4}} \left \vert \int \ du \exp \left(-\frac{1}{2}u.(\frac{\1}{2}-iA)u +iu.(J+i\1)\frac{z_{t}}{2}
\right)\right\vert$$

It is easy to check that the $2n \times 2n$ matrix $\frac{\1}{2}-iA$ obeys
$$\Re (\frac{\1}{2}-iA)>0$$
so that by the formula of Fourier Transform of Gaussian functions, this yields:

\beq
\label{4.7}
\left\vert \det (\frac{\1}{2}-iA)\right\vert^{-1/2}\vert \det (\1-F_{t})\vert^{-1/2}\left \vert\exp \left(- \frac{z_{t}^2}{4}- w.(\1 -2iA)^{-1}w\right)
\right\vert
\eeq
where $w $ denotes the following {\bf complex} vector in $\mathbb C^{2n}$:
$$w:= \frac{1}{2}(J+i\1)z_{t} $$
Thus (\ref{4.7}) can be rewrtitten as
$$\left \vert \det (\frac{\1}{2}-iA)(\1-F_{t})\right \vert^{-1/2} \left \vert \exp\left(-\frac{1}{4}z_{t}.(\1 +K)z_{t}\right)\right\vert$$
where we have defined $K$ as in equ. (3.24)

Recollecting the various prefactors, and reintroducing the parameter $\hbar$, we finally get as a semiclassical approximant of $R(\alpha, t)$ the following expression:
\beq
\label{4.8}
R(\alpha, t)\simeq \left\vert \det (\1 - F_{t}) (\frac{\1}{2}-iA)\right\vert^{-1/2}\left \vert \exp\left(-\frac{1}{4\hbar}z_{t}.(\1 +K)z_{t}\right)
\right\vert
\eeq
We have the following Lemma:

\begin{lemma}
i) We have:
$$\frac{\1}{2}-iA = N(F-\1)^{-1}$$
where N is a $2n \times 2n$ matrix such that $\vert\det N \vert\ge 1$\\

ii)The matrix $K$ is such that:
$$\vert\exp (z.Kz)\vert \le \exp(z^2)$$
for all $z \in \mathbb R^{2n}$.
\end{lemma}

Proof:\\

$\frac{\1}{2} -iA = -\frac{iJ}{2}\left( -iJ(F- \1)+ F+\1\right)(F-\1)^{-1} = N (F-\1)^{-1}$ where 
\beq
\label{4.9}
N := - \frac{iJ}{2}\left((\1 -iJ)F+ \1 + iJ\right)
\eeq

which has the 4-block form:
\beq
\label{4.10}
N= -\frac{iJ}{2}\left(
\begin{array}{cc}
\1 +Z & Z' +i\1\\
i(Z-\1) & i(Z'- i\1)
\end{array}
\right)
\eeq

where $Z = a -ic, \quad Z' = b-id$ and $a,\ b,\ c,\ d$ are the $n \times n$ block matrices of which F is composed:
$$F = \left(
\begin{array}{cc}
a & b\\
c & d
\end{array}
\right)$$
But using the symplecticity of F, the result that $\vert\det N \vert\geq 1$ can be easily checked.\\

ii) is established in \cite{coro3}. It uses the fact that $w.w=0$ and that $\frac{\1}{2}-iA$ has all eigenvalues located on the circle $\vert z-1 \vert =1$, with origin
excluded.
\sq\\

Then we conclude that the prefactor in (\ref{4.8}) is $c_{t}:= \vert \det N \vert ^{-1/2}\le 1$, which completes the proof of Theorem  3.6.
\sq\\

\section{CONCLUDING REMARKS}

Our work is a first mathematical approach of estimates for the Quantum Return Probabilities 
(and Quantum Revivals), and of the Quantum Fidelity for a rather large class 
of Hamiltonians. It also contains a rigorous proof of the Mehlig-Wilkinson formula for the 
Metaplectic Transformations, which appear to be an useful tool, when combined
with the coherent states framework, to understand the semiclassical behavior of the Return
 Probability. Then we are able to describe under which conditions on the linearized 
flow we can have Quantum Revivals for wavefunctions being Coherent States located on a 
{\bf periodic orbit} of the classical flow. We stress that a similar tratment can be 
performed for a semiclassical analysis of the Quantum Fidelity. This will be developed 
elsewhere \cite{coro3}.\\

Here we have concentrated our attention on the Linear Response Regime for the Quantum 
Fidelity. It has been already studied in the physics literature by several authors (\cite
{ceto2},\cite{pro}, \cite{prozni}, \cite{vahe1}). Furthermore the link of the Fidelity 
Decay with the question of Decoherence has been put forward (\cite{cupawi}, 
\cite{cudapazu}, \cite{fihe}, \cite{japa1}, \cite{japa2}, \cite{prose}, \cite{znipro}). It is an interesting topic that deserves further investigation, but that we have not
worked upon.\\

Since no complete review is up to now available on Quantum Fidelity, contrarily to the
 Quantum Revivals problem \cite{ro}, we have provided in the bibliography, for the 
readers' convenience, a rather long (however not exhaustive!) list of references on the physics 
literature on Quantum Fidelity. It is not the purpose here to discuss them in 
detail. To my opinion, one of the prominent points in these approaches is that they
 interestingly distinguish several regimes (Fermi-Golden-Rule, Semiclassical, Linear 
 Response)
where significantly different time behavior of the fidelity manifest themselves; moreover
 they study the link between these behaviors and the regular/chaotic nature of the
underlying classical motion. We think that these topics open new research lines in 
mathematical physics.\\

Let us stress finally that the results developed in this paper, regarding Quantum Revivals as 
well as Quantum Fidelity, show a delicate interplay between {\bf time t, Planck 
constant, and the size of the perturbation $\lambda$}. In particular in the Linear Response 
Regime, the parameter that governs the fidelity decay is the combination 
$t \lambda/\hbar$. Regarding the Semiclassical Regime, we do not seem up to now to be 
able to go beyond the so-called ``Ehrenfest time'' $C \vert \log \hbar \vert$.

\bigskip
{\bf Acknowledgements}
The author acknowledges useful discussions with X. Artru and D. Robert about the topics 
of this paper. She is also indebted to the anonymous referees for numerous useful 
comments and criticisms on a first draft of this work.

\newpage
\bibliographystyle{amsalpha}

\end{document}